\documentclass[11pt]{article}

\usepackage{graphics}

\newcommand{\beq}{\begin{equation}}
\newcommand{\eeq}{\end{equation}}
\newcommand{\beqa}{\begin{eqnarray}}
\newcommand{\eeqa}{\end{eqnarray}}

\renewcommand{\theequation}{\thesection.\arabic{equation}}

\newcommand{\remove}[1]{}
\newcommand{\tcii}{${\rm TC2}$ }

\def\st{\sin\theta}
\def\ct{\cos\theta}
\def\sp{\sin\phi}
\def\cp{\cos\phi}
\def\ttphi{\tan^2\phi}
\def\ssphi{\sec^2\phi}
\def\ccphi{\csc^2\phi\sec^2\phi}
  
\def\gp{g^\prime}
\def\gq{g_l}
\def\gl{g_h}

\begin{document}

\begin{titlepage}

\title{Electroweak Limits on Non-Universal Z' Bosons}

\author{R.S. Chivukula\thanks{e-mail address: sekhar@bu.edu} $^1$ 
and E.H. Simmons\thanks{e-mail address: simmons@bu.edu} $^{1,2}$ \\
$^1$ Department of Physics, Boston University, \\ 590 Commonwealth Ave., 
Boston MA  02215 \\
$^2$ Physics Department, Harvard University, Cambridge MA, 02138}
 
\maketitle

\thispagestyle{empty}

  \begin{picture}(0,0)(0,0)
    \put(400,230){BUHEP-02-16}
    \put(400,210){HUPT-02/A002}
  \end{picture}
  \vspace{24pt}
  

\begin{abstract}

Many types of physics beyond the standard model include an extended
electroweak gauge group. If these extensions are associated with
flavor symmetry breaking, the gauge interactions will not be
flavor-universal. In this note we update the bounds placed by
electroweak data on the existence of flavor non-universal extensions
to the standard model in the context of topcolor assisted technicolor
(TC2), noncommuting extended technicolor (NCETC), and the ununified
standard model (UUM). In the first two cases the extended gauge
interactions couple to the third generation fermions differently than
to the light fermions, while in the ununified standard model the gauge
interactions couple differently to quarks and leptons.  The extra
$SU(2)$ triplet of gauge bosons in NCETC and UUM models must be
heavier than about 3 TeV, while the extra $Z$ boson in TC2 models must
be heavier than about 1 TeV.
\end{abstract}

\end{titlepage}

\newpage
\renewcommand{\thepage}{\arabic{page}}
\setcounter{page}{1}

\section{Introduction}
\label{sec:intro}
\setcounter{equation}{0}

Precision electroweak data place bounds on possible extensions of the
electroweak gauge group. If these extensions are associated with
flavor symmetry breaking, the gauge interactions will not be
flavor-universal \cite{Li:1981nk,Georgi:1989ic,Hill:1994hp,
Chivukula:1994mn,Muller:1996dj,Malkawi:1996fs}. Three
particular models which contain such flavor non-universal gauge
interactions are topcolor assisted technicolor (TC2)
\cite{Hill:1994hp}, noncommuting extended technicolor
(NCETC)\cite{Chivukula:1994mn}, and the ununified standard model
(UUM)\cite{Georgi:1989ic,Georgi:1989xz}. In the first two cases the
extended gauge interactions couple to the third generation fermions
differently than to the light fermions, while in the ununified standard model
the gauge interactions couple differently to quarks and leptons.

In this note we update the bounds
\cite{Chivukula:1996cc,Chivukula:1996gu,Chivukula:1994qw} placed by
precision data on the existence of the extra neutral $Z$ boson in TC2
models and the extra $SU(2)$ triplet of gauge bosons in NCETC and UUM
models. We also consider the bounds arising from the search for
contact interactions in scattering experiments, and from requiring the
absence of large flavor-changing neutral currents\footnote{Related
precision electroweak and flavor-changing neutral current bounds on
generic Z' bosons may be found in \protect\cite{langacker}.} (for
CKM-like values of the various mixing angles).

We find that the the extra $Z$ in TC2 models must be heavier than
about 2 TeV for generic values of the gauge coupling.  However, the
TC2 limits illustrate that, for specific values of the parameters,
cancellations can limit the size of deviations of Z-pole observables
-- allowing for a $Z'$ as light as 730 GeV. In such a case, off-shell
measurements become important.  Specifically, limits on contact
interactions at LEPII imply that the TC2 $Z'$ must be heavier than
about 1 TeV.  Note that these limits hold regardless of the assumed
flavor structure of the quark mixing matrices -- unlike the
potentially stronger but more model-dependent limits from B-meson
mixing.

In contrast, the extra $SU(2)$ triplet of gauge bosons in NCETC
and UUM models must be somewhat heavier, with masses always greater
than about 3 TeV.  For these models, the limits from Z-pole
observables are stronger than those from contact interactions at LEP
II or from flavor-changing neutral currents.

\section{Z' Bosons in TC2 Models}
\label{sec:tc2}
\setcounter{equation}{0}
\subsection{TC2 models}

In technicolor models \cite{technicolor}, electroweak symmetry
breaking occurs when a new asymptotically free gauge theory
(technicolor) spontaneously breaks the chiral symmetries of the new
fermions to which it couples (technifermions).  Small fermion masses
can be generated if the technicolor group is embedded in a larger
extended technicolor (ETC) gauge interaction felt by ordinary and
technifermions alike \cite{extendedtc}.  The key feature of
topcolor-assisted technicolor models \cite{Hill:1994hp} is that an
additional, larger, component of the top quark mass is dynamically
generated by extended color interactions (topcolor \cite{Hill:1991at})
at a scale of order 1 TeV.  The topcolor interactions may be flavor
non-universal (as in classic TC2 \cite{Hill:1994hp}) or
flavor-universal \cite{Popovic:1998vb}.  In either case, a
non-universal extended hypercharge group is often invoked
\cite{Hill:1994hp,Buchalla:1995dp} to ensure that the top quark
condenses and receives a large mass while the bottom quark does not.

The electroweak gauge symmetry in \tcii models is therefore $SU(2)_L
\times U(1)_1 \times U(1)_2$. Here $U(1)_1$ is a weak gauge
interaction and $U(1)_2$ is the, presumably strong, interaction with
isospin-violating quark couplings that facilitates top-quark, but not
bottom-quark, condensation.  The required pattern of electroweak
gauge symmetry breaking is more complicated than that in ordinary
technicolor models; it generally involves two scales (rather than just
one) to break the $SU(2)_L \times U(1)_1 \times U(1)_2$ symmetry down
to $U(1)_{em}$.  The required pattern of breaking is:
\begin{center}
$SU(2)_{L} \otimes U(1)_1 \otimes U(1)_2 $
\end{center}
\vspace{-20pt}
\begin{center}
$\downarrow \ \ \ \ \  u $
\end{center}
\vspace{-20pt}
\begin{center}
$SU(2)_{L} \otimes U(1)_Y$
\end{center}
\vspace{-20pt}
\begin{center}
$\downarrow\ \ \ \ \ v $
\end{center}
\vspace{-20pt}
\begin{center}
$U(1)_{em}$,
\end{center}
where hypercharge, $Y=Y_1 + Y_2$, is equal to the sum of the generators
of the two $U(1)$'s.

The gauge couplings may be written
\beq
g ={e\over \st}\,,\quad
g_1' = {g^\prime\over\cp} = {e\over \cp\ct}\,,\quad
g_2' = {g^\prime\over\sp} = {e\over \sp\ct}\,,
\eeq
in terms of the usual weak mixing angle $\theta$ and a new mixing angle
$\phi$.
It is convenient to rewrite the neutral gauge bosons in terms of the photon,
\beq
A^\mu = \ct\,(\cp\, B^\mu_1 + \sp\, B^\mu_2) + \st\, W^\mu_3,
\eeq
which couples to electric charge $Q$ with strength $e$, a field
\beq
Z^\mu_1=-\st\,(\cp\, B^\mu_1 + \sp\, B^\mu_2)+\ct\, W^\mu_3,
\eeq
which couples as the standard model $Z$ would couple,
to $T_3 - Q \sin^2\theta$  with strength
${e\over \st\ct}$
and the field
\beq
Z^\mu_2 = -\sp\, B^\mu_1 + \cp\, B^\mu_2,
\label{tctzpfir}
\eeq
which couples to the current $Y^\prime = Y_2 - \sin^2\phi Y$ with
strength ${e\over \ct\sp\cp}$.
In this basis, using the relation $Q=T_3 +Y$
and the fact that $Q$ is conserved, the
mass-squared matrix for the $Z_1$ and $Z_2$ can be written as:
\beq
M_Z^2=\left({e  \over {2 \st \ct}} \right)^2\,
\pmatrix{<T_3 T_3>&
{ {\sin \theta }\over {\sp \cp}} <T_3 Y^\prime > \cr
{ {\sin \theta }\over {\sp \cp}} <T_3 Y^\prime >&
{{\sin^2 \theta}\over {\sin^2\phi \cos^2\phi}} <Y^\prime Y^\prime>\cr}\,,
\eeq
where, from the charged-$W$ masses we see that
$<T_3 T_3> = v^2 \approx (250\, {\rm GeV})^2$.

As discussed in \cite{Chivukula:1996cc}, in natural \tcii models
\cite{Lane:1995gw} the expectation value leading to
$Z_1 - Z_2$ mixing, \nobreak{$<T_3 Y^\prime>$}, can be calculated {\it
entirely} in terms of the gauge couplings, $v$, and the $Y_2$ charges
of the left- and right-handed top quark.  Using the definition of
$Y^\prime$, we see that 
\beq 
<T_3 Y^\prime> \, = \, <T_3 Y_2> - \sin^2\phi <T_3 Y>\, .  
\eeq 
Since $Y=Q-T_3$ and $Q$ is conserved, the
last term is equal to $+\sin^2\phi <T_3 T_3>$. Furthermore in natural
\tcii models, since the technifermion $Y_2$-charges are assumed to be
isospin symmetric, the technifermions do not contribute to the first
term. The only contribution to the first term comes from the top-quark
condensate 
\beq 
{<T_3 Y_2> \over <T_3 T_3>} =
2(Y^{t_L}_2-Y^{t_R}_2){f^2_t \over v^2}\, , 
\eeq 
where $f_t$ is the
analog of $f_\pi$ for the top-condensate and is equal to \cite{F} 
\beq
f_t^2 \approx {{N_c}\over{8\pi^2}}\, m_t^2
\log\left({{M^2}\over{m_t^2}}\right) 
\eeq 
in the Nambu---Jona-Lasinio
\cite{NJL} approximation, and $M$ is the mass of the extra color-octet
gauge bosons (colorons) arising in the extended color
interactions. For $m_t \approx 175$ GeV and $M\approx 1$ TeV, we find
$f_t \approx 64$ GeV.

If we define
\beq
x \equiv {{\sin^2 \theta}\over {\sin^2\phi \cos^2\phi}}  \,
{{<Y^\prime Y^\prime>}\over{ <T_3 T_3>}}  \, \propto {u^2\over v^2} \, ,
\eeq
and
\beq
\epsilon \equiv   2  \,  {{f_t^2}\over{v^2}}
\left( Y_2^{t_L} - Y_2^{t_R} \right)\, ,
\eeq
the $Z_1 - Z_2$ mass matrix can be written as
\beq
M_Z^2=M^2_{Z}|_{\rm SM}\,
\pmatrix{
1 & {\tan\phi\sin \theta }\left(1 +{\epsilon\over\sin^2\phi} \right) \cr
{\tan\phi \sin \theta}  \left( 1 +{\epsilon\over\sin^2\phi}\right)  & x \cr}
\, .
\label{tctzpms}
\eeq
In the large-$x$ limit the mass eigenstates are
\beqa
Z  & \approx & Z_1 - {{\tan\phi \sin \theta }\over {x}} \left( 1 +
{{\epsilon}\over{\sin^2\phi}} \right)Z_2 \\
Z^\prime & \approx & {{\tan\phi \sin\theta }\over {x}} \left( 1 +
{{\epsilon}\over{\sin^2\phi}} \right) Z_1 + Z_2
\label{tctzpeig}
\eeqa
The shifts in the $Z$ coupling to $f {\overline f}$ (with $e/(\cos\theta
\sin\theta)$ factored
out) are
therefore given by:
\beq
\delta g^f \approx -  \, {{\sin^2 \theta}\over{x \cos^2\phi}}  \, \left( 1 +
{{\epsilon}\over{\sin^2\phi}}\right)
\left[Y_2^f -\sin^2\phi Y^f \right]\, .
\label{dg}
\eeq
Mixing also shifts the $Z$ mass, and gives a contribution to the
$T$ parameter \cite{e:pestak} equal to:
\beq
\alpha T \approx { {\tan^2\phi \,\sin^2 \theta}\over {x}} \left(1 +
{{\epsilon}\over{\sin^2\phi}}\right)^2\, .
\label{T}
\eeq

The shifts in the $Z$-couplings and mass are sufficient to describe
electroweak phenomenology on the $Z$-peak. For low-energy processes,
in addition to these effects we must also consider the effects of
$Z^\prime$-exchange. To leading order in $1/x$, these effects may
be summarized by the four-fermion interaction \cite{Chivukula:1996cc}
\beq
-{\cal L}^{Z^\prime}_{\rm NC} = {4 G_F \over \sqrt{2}}  \,
{\sin^2\theta \over x \sin^2\phi \cos^2\phi}  \,
\left(J_{Y_2} - \sin^2\phi J_Y \right)^2\, ,
\eeq
where $J_{Y_2}$ and $J_Y$ are the $Y_2$- and hypercharge-currents,
respectively.  It is useful to note that if $\epsilon$ is negative, then
all the $Z$ pole mixing effects (equations (\ref{dg}) and (\ref{T}))
vanish when $\sin^2\phi = -\epsilon$, although the low-energy effects of
$Z^\prime$ exchange do not.

An important consistency check is whether the Landau pole of the
strongly-coupled $U(1)_2$ gauge interaction lies sufficiently
far above the symmetry-breaking scale to render the theory
self-consistent.  In \cite{Popovic:1998vb}, it was shown that a factor
of 10 separation of scales is ensured for $\kappa_1 < 1$ where
\begin{equation}
\kappa_1 \equiv \frac{\alpha_{em}}{\cos^2\theta_W} 
\left( \frac{g_2}{g_1} \right)^2
\end{equation}
and $g_1$ ($g_2$) is the coupling of the $U(1)$ group under which the
first and second (third) generation fermions are charged.  Since the
ratio of coupling constants is defined to be the cotangent of the
gauge boson mixing angle $\phi$, the constraint on $\kappa_1$ will be
met if
\begin{equation}
\sin^2\phi > \left[ 1 + \frac{\cos^2\theta_W}{\alpha_{em}}
\right]^{-1} \approx 0.01\ \ .
\end{equation}
This condition is satisfied for the values of $\sin^2\phi$ considered
in our analysis.

\subsection{Precision EW Constraints}

\begin{table}[bth]
\begin{center}
\begin{tabular}{|c||l|l||c|c|c|c|}\hline\hline
Quantity & Experiment & SM & TC2 & NCETC & NCETC & UUM \\
& & & & (heavy) & (light) & \\
\hline \hline
$\Gamma_Z$   & 2.4952 $\pm$ 0.0023 & 2.4962  & * & * & * & * \\
$A_{LR}$ &   0.1514 $\pm$ 0.0022 & 0.1482    & * & * & * & * \\
$A_{FB}^e$ &   0.0145 $\pm$ 0.0025 & 0.0165  & * & * & * &   \\
$A_{FB}^\mu$ & 0.0169 $\pm$ 0.0013 & 0.0165  & * & * & * &   \\
$A_{FB}^\tau$ & 0.0188 $\pm$ 0.0017 & 0.0165 & * & * & * &   \\
$A_{FB}^\ell$ & 0.0171 $\pm$ 0.0010 & 0.0165 &   &   &   & * \\
$\sigma_h^{non}$ 
           & 41.541 $\pm$ 0.037 & 41.480     & * & * & * &   \\
$\sigma_h^{univ}$ 
           & 41.540 $\pm$ 0.033 & 41.481     &   &   &   & * \\
$R_b$ & 0.21646 $\pm$ 0.00065 & 0.215768     & * & * & * & * \\
$R_c$ & 0.1719 $\pm$ 0.0031 & 0.1723         & * & * & * & * \\
$R_e$ & 20.804 $\pm$ 0.050 & 20.741          & * & * & * &   \\
$R_\mu$ & 20.785 $\pm$ 0.033 & 20.741        & * & * & * &   \\
$R_\tau$ & 20.764 $\pm$ 0.045 & 20.741       & * & * & * &   \\
$R_\ell$ & 20.767 $\pm$ 0.025 & 20.741       &   &   &   & * \\
$A_{e}(P_\tau)$ & 0.1498 $\pm$0.0049 & 0.1482& * & * & * &   \\
$A_{\tau}(P_\tau)$ &0.1439 $\pm$0.0043&0.1482& * & * & * &   \\
$A_{\ell}(P_\tau)$ & 0.1465 $\pm$0.0033&0.1482&   &   &   & * \\
$A_{FB}^b$ & 0.0994 $\pm$ 0.0017 & 0.1039    & * & * & * & * \\
$A_{FB}^c$ & 0.0707 $\pm$ 0.0034 & 0.0743    & * & * & * & * \\
${\cal A}_b$ & 0.922 $\pm$ 0.020 & 0.935     & * & * & * & * \\
${\cal A}_c$ & 0.670 $\pm$ 0.026 & 0.668     & * & * & * & * \\
$M_W$ (LEP II) & 80.450 $\pm$ 0.039 & 80.394 & * & * & * & * \\
$M_W$ (Tevatron) & 80.454 $\pm$ 0.060 &80.394& * & * & * & * \\
$g_L^2$ & 0.3005 $\pm$ 0.0014 & 0.3042       & * & * & * & * \\
$g_R^2$ & 0.0310 $\pm$ 0.0011 & 0.0301       & * & * & * & * \\
$Q_W(Cs)$ & -72.39 $\pm$ 0.59 & -72.89        & * & * & * & * \\
$A_e$ & 1.0012 $\pm$ 0.0053  & 1.0  &   & * &   &   \\
\hline\hline
\end{tabular}
\end{center}
\caption[extab]{Experimental and predicted SM values of electroweak
observables.  Experimental values of most quantities are from
\protect\cite{Abbaneo:2002xx}; the experimental value of $M_W$(Tevatron)
and of $A_e$, the ratio of $G^2_F$ as inferred from the decays of
$\tau \to e$ vs. $\mu \to e$, are from \cite{pdbook}; the experimental
values of $g^2_L$ and $g^2_R$ are from \protect\cite{Zeller:2001hh}.
The theoretical SM values are from \protect\cite{Abbaneo:2002xx}. 
The value of $\sigma_h$ labeled ```univ'' (``non'') was derived 
(without) assuming lepton universality.  In
each of the last four columns, a * indicates that the predicted value
of the observable in the relevant model differs from that in the SM
(see Appendices); thus, 22 observables were used in the TC2 and light
ETC fits; 23 in the heavy ETC fit, and 17 in the UUM fit. }
\label{tab:obs-sm}
\end{table}

In the presence of the enlarged electroweak gauge group in TC2 models,
the predicted properties of the $Z^0$ resonance and of the low-energy
weak interactions are altered.  Quantities affected at the Z pole
include the width ($\Gamma_Z$), decay asymmetries ($A_{LR}$,
$A_{FB}^e$, $A_{FB}^\mu$, $A_{FB}^\tau$, $A_e(P_\tau)$,
$A_\tau(P_\tau)$, $A_{FB}^b$, $A_{FB}^c$, ${\cal A}_b$, ${\cal A}_c$),
peak hadronic cross-section ($\sigma_h$), and partial-width ratios
($R_b$, $R_c$, $R_e$, $R_\mu$, $R_\tau$).  Other affected observables
are the W mass, the rates of deep-inelastic neutrino-nucleon
scattering ($g_L^2$, $g_R^2$) and the degree of atomic parity
violation ($Q_W(Cs)$).  We have used the general approach of
ref. \cite{Burgess:1993vc} to calculate how the presence of the $Z'$
modifies the predicted values of the electroweak observables whose
measured and SM values are listed in Table \ref{tab:obs-sm}.  The
formulae for these leading (tree-level) alterations are presented in
Appendix A as functions of the mixing angle, $\phi$, and the ratio of
squared vevs, $1/x$.

\begin{figure}[thb]
\begin{center}
{\rotatebox{0}{\scalebox{1.0}{\includegraphics{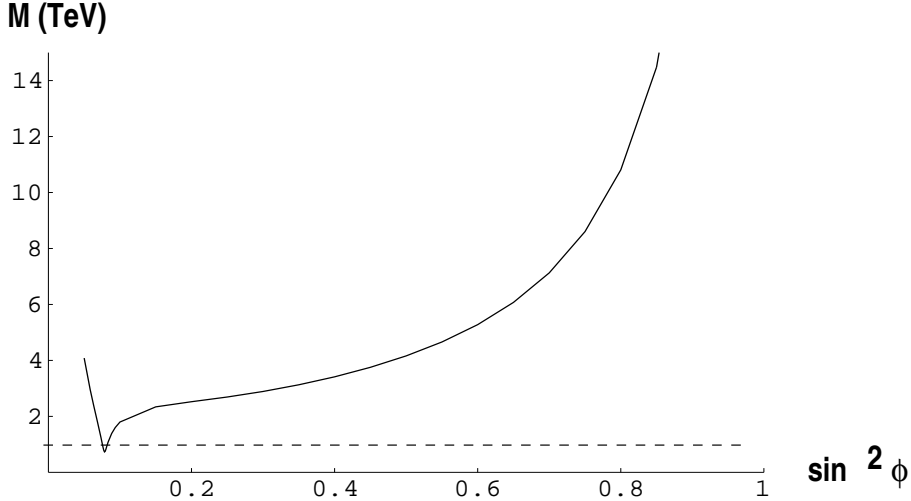}}}}
\end{center}
\caption[down]{\small Lower bounds on the Z' boson mass 
at 95\% CL in TC2 models
as a function of mixing angle.  The solid curve is the lower bound from
precision electroweak data; the dashed line is the lower bound from
LEP II contact interaction studies.}
\label{fig:tc2}
\end{figure}

We have performed global fits of the electroweak data to the
expressions in Appendix A, allowing $1/x$ and $\phi$ to vary.  More
precisely, at each value of $\phi$ we determined a best-fit value of
$1/x$, along with one-sigma errors, and used the relation between
$1/x$ and $M_{Z'}$ from eqn. (\ref{tctzpms}) to translate that into a
95\% c.l. lower bound\footnote{The standard deviation on which the
confidence level is based takes the number of free parameters into
account as well as the number of measured quantities.} on $M_{Z'}$.
Figure \ref{fig:tc2} summarizes these results.  We find that the mass
of the Z' boson can be below 1 TeV for 0.0744 $\leq \sin^2\phi \leq$
0.0834, with a minimum value of about 730 GeV.  This is more than a
factor of two tighter than the bound set in
ref. \cite{Chivukula:1996cc}.  The goodness-of-fit\footnote{The
goodness-of-fit expresses the likelihood that the measurements would
give a $\chi^2$/d.o.f. this large if the model is correct. Again, the
number of d.o.f. equals the number of measurements minus the number of
free parameters.}  for the TC2 model with the Z' lying on the
lower-bound curve is 4.2\%, somewhat lower than the 5.6\% result when we
fit the SM predictions to the same data.

\begin{figure}[hbt]
\begin{center}
{\rotatebox{0}{\scalebox{1.0}{\includegraphics{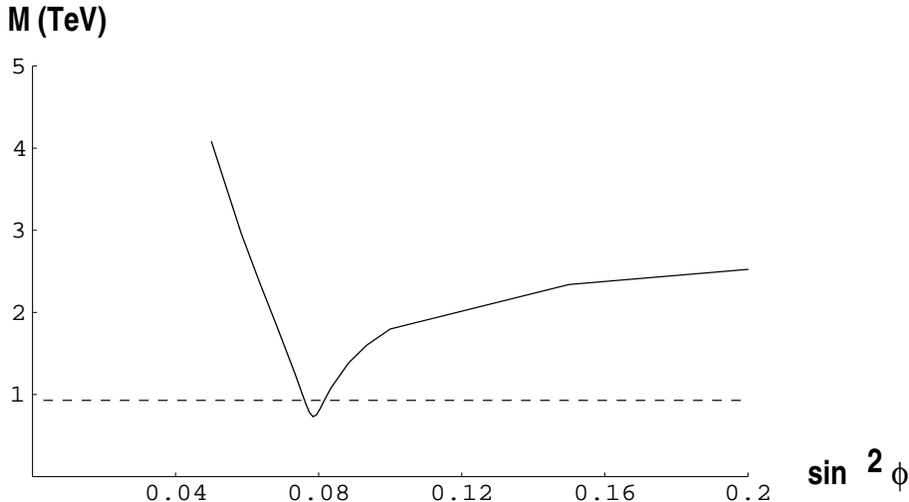}}}}
\end{center}
\caption[down]{\small Close-up view of lower bounds on TC2 Z' boson
mass at 95\% CL as a function of mixing angle. Solid curve is lower bound from
precision electroweak data; dashed line is lower bound from LEP II
contact interaction studies.}
\label{fig:tc2f}
\end{figure}

\subsection{Contact Interactions at LEP II}

 The LEP experiments have recently published limits on contact
interactions \cite{Abbaneo:2002xx} which may be used to set a lower
bound\footnote{Limits have also been set by the Tevatron experiments,
but since these involve only fermions of the first and second
generations, the expression for $M_{Z'}$ analogous to the RHS of
eqn. (\ref{zplime}) is suppressed by a factor of $\cos\phi / \sin\phi$
which renders the associated bound weaker than that from LEP data
\cite{Lynch:2001md}.} on $M_{Z'}$.  Following the notation of
\cite{Eichten:1983hw}, they write the effective Lagrangian for the
four-fermion contact interaction in the process $e^+ e^- \rightarrow f
\bar{f}$ as
\begin{equation} 
\mathcal{L}_{\rm{contact}} = \frac{g^2}{\Lambda^2 (1 + \delta)}
\sum_{i,j = L, R} \eta_{ij} \big( \bar{e}_i \gamma_\mu
e_i\big) \big(\bar{f}_j \gamma^\mu f_j\big)
\label{lepcontact}
\end{equation}
where $\delta = 1$ if $f$ is an electron and $\delta = 0$ otherwise.
The values of the coefficients $\eta_{ij}$ set the chirality structure
of the interaction being studied; the LEP analysis always takes one of
the $\eta_{ij}$ equal to 1 and sets the others to zero.  Following the
convention \cite{Eichten:1983hw} of taking $g^2 / 4\pi = 1$, they
determine a lower bound on the scale $\Lambda$ associated with each
type of new physics.  In fact, they determine separate limits
$\Lambda^+$ and $\Lambda^-$ for each case, depending on whether
constructive or destructive interference is assumed.  Of particular
interest to us for TC2 models are their limits on contact interactions
where the final-state fermions are the third-generation fermions
$\tau$ or $b$.

At energies well below the mass of the TC2 Z' boson, its exchange in the
process $e^+ e^- \to f \bar{f}$ where $f$ is a $\tau$ lepton or
b-quark may be approximated by the contact interaction
\begin{equation} 
{\cal{L}}_{NC} \supset \frac{e^2}{\cos^2\theta\, M^2_{Z'}}
\left(\frac{c_\phi}{s_\phi} Y_{e_i} \big(\bar{e_i} \gamma_\mu
  e_i\big)\right) \left(\frac{s_\phi}{c_\phi} Y_{f_j}
  \big(\bar{f}_j \gamma^\mu f_j\big)\right)\ , 
\end{equation}
based on the Z'-fermion couplings implied by
eqns.~(\ref{tctzpfir},\ref{tctzpeig}).  Comparing this with the contact
interactions studied by LEP (\ref{lepcontact}), we find
\begin{equation}
M_{Z'} = \Lambda^{sgn[Y_{e_i} Y_{f_j}]} \sqrt{\frac{\alpha_{em}}{
\cos^2\theta }|Y_{e_i} Y_{f_j}|}\ .
\label{zplime}
\end{equation}
Thus, when the produced fermions are tau leptons or right-handed
$\tau$-leptons, the LEP limit on $\Lambda^+$ is the relevant one; when
left-handed b-quarks are produced, the limit on $\Lambda^-$ rules.

By using the LEP limits on contact interactions in equation
(\ref{zplime}), we find that the strongest lower bound on $M_{Z'}$ comes
from production of right-handed tau leptons.  LEP sets the limit
\cite{Abbaneo:2002xx} $\Lambda^+_{RR} \geq 10.9$ TeV.  This translates to
the lower bound $M_{Z'} \geq 1.09$ TeV, independent of the mixing
angle $\phi$.  Comparing this with the bounds from precision
electroweak data derived in the previous subsection, we see that the
region of lower Z' mass (down to 730 GeV) previously allowed at
$\sin\phi \approx 0.0784$ is now eliminated.

\subsection{Contrasting Limits from B-meson mixing}

Recent work in the
literature \cite{Kominis:1995fj,Burdman:2001in,Hill:1994hp} has shown
that lower bounds on the mass of the Z' boson in TC2 models may be
extracted from limits on neutral B-meson mixing.  These limits turn
out to be quite sensitive to the flavor structure of the model.  For
example, ref.  \cite{Burdman:2001in} shows that in classic TC2 models
\cite{Hill:1994hp} in which all quark mixing is confined to the
left-handed down-quark sector, one must have $M_{Z'} > 6.8$ TeV (9.6
TeV) if ETC does (does not) contribute to the Kaon CP-violation
parameter $\epsilon$.  This is a stricter limit than we have found
over much of the parameter space of the model.
Ref. \cite{Simmons:2001va} shows that in flavor-universal TC2 models
\cite{Popovic:1998vb}, if one makes the same assumption about the
flavor structure, the corresponding lower limit on the Z' mass is
merely 590 GeV (910 GeV) -- that is, weaker than our bounds.
Moreover, changing the assumed flavor structure alters the implied
limits.  In contrast, our electroweak and contact-interaction limits
hold for all models with the gauge and fermion sector described at the
start of this section.

\section{Weak Bosons in NCETC}
\label{sec:ncetc}
\setcounter{equation}{0}

\begin{figure}
\begin{center}
{\rotatebox{0}{\scalebox{1.0}{\includegraphics{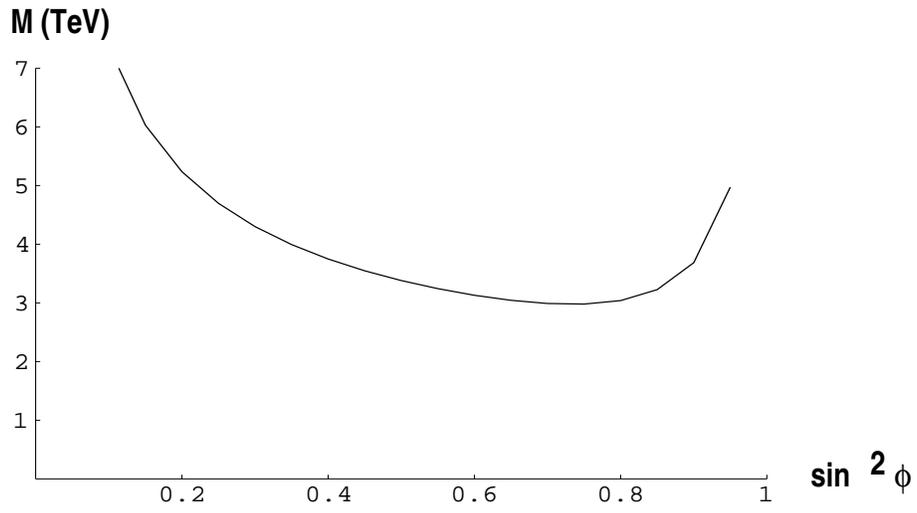}}}}
\end{center}
\caption[dow]{\small Lower bound on heavy NCETC Z' boson mass 
at 95\% CL as a function of
mixing angle. }
\label{fig:hncetc}
\end{figure}

\begin{figure}
\begin{center}
{\rotatebox{0}{\scalebox{1.0}{\includegraphics{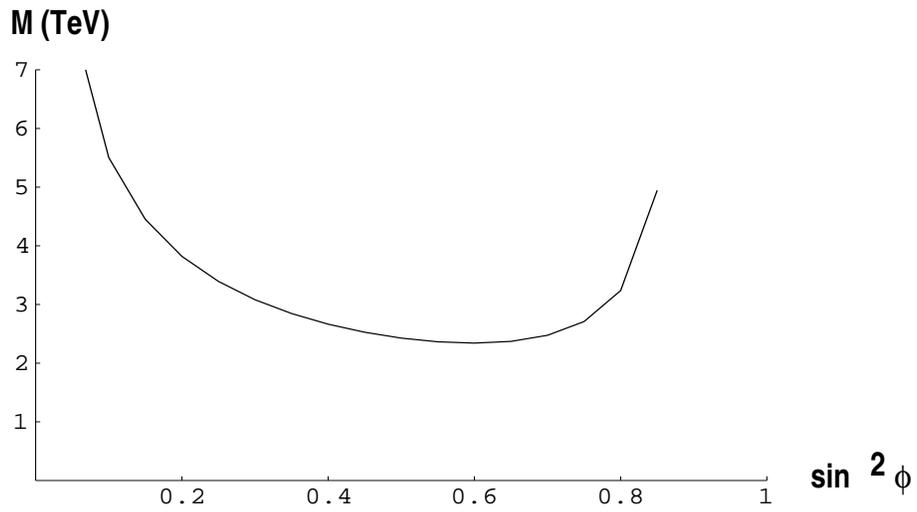}}}}
\end{center}
\caption[do]{\small Lower bound on light NCETC Z' boson mass at 95\%
CL as a function of
mixing angle. }
\label{fig:lncetc}
\end{figure}

In extended technicolor models \cite{extendedtc}, fermion masses are
generated because the ETC gauge bosons couple the ordinary quarks and
leptons to the technifermion condensate.  The large mass of the top
quark arises through ETC dynamics at a relatively low scale, not far
above the scale of electroweak symmetry breaking.  The defining
characteristic of non-commuting extended technicolor (ETC) models
\cite{Chivukula:1994mn, Chivukula:1996gu} is that the ETC interactions
do not commute with the $SU(2)_L$ interactions of the standard model.
That is, the weak interactions are partially embedded in the ETC gauge
group.  Providing masses for one family of ordinary fermions (say, the
third family) then requires a pattern of gauge symmetry breaking with
three distinct scales:
\begin{center}
$G_{ETC}  \otimes SU(2)_{light} \otimes U(1)' $
\end{center}
\vspace{-20pt}
\begin{center}
$\downarrow \ \ \ \ \ f $
\end{center}
\vspace{-20pt}
\begin{center}
$G_{TC} \otimes SU(2)_{heavy}  \otimes SU(2)_{light} \otimes U(1)_Y $
\end{center}
\vspace{-20pt}
\begin{center}
$\downarrow\ \ \ \ \ u $
\end{center}
\vspace{-20pt}
\begin{center}
$G_{TC}  \otimes SU(2)_{L} \otimes U(1)_Y$
\end{center}
\vspace{-20pt}
\begin{center}
$\downarrow\ \ \ \ \ v $
\end{center}
\vspace{-20pt}
\begin{center}
$G_{TC}  \otimes U(1)_{em}$,
\end{center}
The ETC gauge group is broken to technicolor and an
$SU(2)_{heavy}$ subgroup at the scale $f$.  The $SU(2)_{heavy}$ gauge
group is effectively the weak gauge group for the third 
generation\footnote{Experimental limits on the heavy gauge bosons of 
topflavor models \protect\cite{Muller:1996dj,Malkawi:1996fs} which 
have an identical electroweak gauge structure but use fundamental 
higgs bosons to effect mass generation are considered in 
\protect\cite{Lee:1998qq,Malkawi:1999sa}.} in
these non-commuting ETC models, while the $SU(2)_{light}$ is the weak
gauge group for the two light generations.  The two $SU(2)$'s are
mixed (i.e. they break down to a diagonal $SU(2)_L$ subgroup) at the
scale $u$.  Finally electroweak symmetry breaking
is accomplished at the scale $v$, as is standard in technicolor
theories.

The two simplest possibilities for the $SU(2)_{heavy} \times
SU(2)_{light}$ transformation properties of the order
parameters that produce the correct combination of mixing and
breaking of these gauge groups are:
\beq
\langle \varphi \rangle \sim (2,1)_{1/2},\ \ \ \ \langle
\sigma\rangle \sim (2,2)_0 ~,\ \ \ \ \ \ \ ``{\rm heavy\ case}"~,
\eeq
and
\beq
\langle \varphi \rangle \sim (1,2)_{1/2},\ \ \ \ \langle
\sigma\rangle \sim (2,2)_0 ~,\ \ \ \ \ \ \ ``{\rm light\ case}"~.
\eeq
Here the order parameter $\langle\varphi\rangle$ is responsible for
breaking $SU(2)_L$ while $\langle\sigma\rangle$ mixes
$SU(2)_{heavy}$ with $SU(2)_{light}$.  We refer to these two
possibilities as ``heavy'' and ``light'' according to whether 
$\langle\varphi\rangle$
transforms non-trivially under $SU(2)_{heavy}$ or $SU(2)_{light}$.

The heavy case, in which $\langle\varphi\rangle$ couples to the heavy
group, is the choice made in \cite{Chivukula:1994mn}, and corresponds
to the case in which the technifermion condensation responsible for
providing mass for the third generation of quarks and leptons is also
responsible for the bulk of electroweak symmetry breaking (as measured
by the contribution made to the $W$ and $Z$ masses).  The light case,
in which $\langle\varphi\rangle$ couples to the light group,
corresponds to the opposite scenario: here the physics responsible for
providing mass for the third generation {\it does not} provide the
bulk of electroweak symmetry breaking.  In this respect, the light
case is akin to multiscale technicolor models \cite{multi,chiraltc}
and top-color assisted technicolor \cite{Hill:1994hp}.

The gauge couplings may be written
\beq
g_{light}={e\over \sp \st}\,,\quad
g_{heavy} = {e\over \cp\st}\,,\quad
g' =  {e\over \ct}\,,
\eeq
where $\theta$ is the usual weak angle and $\phi$ specifies the
strength of the additional interactions. Charge is given
by $Q= T_{3l}+T_{3h}+Y~$ and the photon eigenstate, by
\beq
A^\mu = \st \sp \,W_{3l}^\mu + \st \cp \,W_{3h}^\mu +\ct X^\mu~~,
\eeq
where $W_{3l,h}$ are the neutral gauge-bosons in $SU(2)_{light,heavy}$
and $X$ is the gauge-boson of $U(1)_Y$.
It is convenient to discuss the mass eigenstates in the rotated basis
\beqa
W^{\pm}_1 &=& s\,W^{\pm}_l+c\,W^{\pm}_h\,, \\
W^{\pm}_2 &=& c\,W^{\pm}_l-s\,W^{\pm}_h\,, \\
Z_1 &=&\ct\,(s\,W_{3l}+c\,W_{3h})-\st\,X\,, \\
Z_2 &=& c\,W_{3l}-s\,W_{3h}\,,
\eeqa
in which the gauge covariant derivatives separate neatly into standard and
non-standard pieces
\beqa
D^\mu &=&\partial^\mu + ig\left( T_l^\pm + T_h^\pm \right) W^{\pm\,\mu}_1
+ig\left( {c \over s}T_l^\pm - {s \over c}T_h^\pm \right) W^{\pm\,\mu}_2
\nonumber \\
& &  +\  i{g \over {\ct}}\left( T_{3l} + T_{3h} -
\sin^2\theta \,Q \right) Z^\mu_1 +
ig\left( {c \over s}T_{3l} - {s \over c}T_{3h} \right)
Z^\mu_2.
\label{eqn:ncetc-coupl}
\eeqa
where $g\equiv {e \over \st}$.
The breaking of $SU(2)_L$ results in mixing of  $Z_1$ and $Z_2$, as
well as a mixing of $W^{\pm}_1$ and $W^{\pm}_2$.
The mass-squared matrix for the $Z_1$ and $Z_2$ is:
\beqa
M_Z^2&=&\left({e v \over {2 \st}} \right)^2\,
\pmatrix{{1\over \cos^2\theta}&
{- s\over c\cos\theta}\cr{- s\over c\cos\theta}&
{x\over s^2c^2}+{s^2\over c^2}\cr}\,,\ \ \ \ \ \ \ [{\rm heavy\ case}] 
\\
M_Z^2&=&\left({e v \over {2 \st}} \right)^2\,
\pmatrix{{1\over \cos^2\theta}&
{ c\over s\cos\theta}\cr{ c\over s\cos\theta}&
{x\over s^2c^2}+{c^2\over s^2}\cr}\,.\ \ \ \ \ \ \ [{\rm light\ case}]
\eeqa
In these expressions, $x= u^2 / v^2$, and the
mass-squared matrix for $W_1$ and $W_2$ is obtained by setting $\ct =
1$ in the above matrix. 
In the limit of large $x$, the light gauge boson mass eigenstates are
\beqa
W^L &\approx &W_1+{c s^3 \over {x}}\,W_2, \ \ \ \ \ \ 
Z^L\approx Z_1+{c s^3 \over {x\cos\theta}}\,Z_2\
 \ \ \ \ \ \ [{\rm heavy\ case}]
\label{dgzq} \\
W^L&\approx& W_1-{c^3 s \over {x}}\,W_2, \ \ \ \ \ \ 
Z^L\approx Z_1-{c^3 s \over {x\cos\theta}}\,Z_2\,. \
 \ \ \ \ \ \ [{\rm light\ case}]
\label{dmzgfiq}
\eeqa The heavy bosons $W^H$ ($Z^H$) are the orthogonal combinations
of $W_1$ and $W_2$ ($Z_1$ and $Z_2$); their masses are approximately
$M^H_{W,Z} \approx {\sqrt{x}\over sc} M_W^{0}$ where $M^0_W$ is the
tree-level W-boson mass in the Standard Model.

In addition, the extended-technicolor interactions responsible for
giving mass to the third-generation of of quarks and leptons is
expected to give rise to shifts in the couplings to the left-handed
bottom $\delta g^b_L$ and the left-handed leptons $\delta g^\tau_L =
\delta g^{\nu_\tau}_L$.  More precisely, the associated change in the
$Zff$ coupling is of the form $\delta g (e/\st\ct)$.

The presence of the extra electroweak bosons and possible ETC vertex
corrections alters the predicted values of electroweak observables.
The quantities whose values are affected are indicated separately for
the heavy and light cases of NCETC in Table \ref{tab:obs-sm}.  In
light NCETC, the same quantities are affected as in TC2 models; in
heavy NCETC, the value of $A_e$ \cite{pdbook}, the ratio of $G^2_F$ as
inferred from $\tau\to e$ vs. $\mu \to e$, is also altered.
Expressions for the predicted shifts from SM values are discussed in
Appendix B.

We performed separate global fits of the electroweak data to the
parameters $\delta g_b$, $\delta g_\tau \equiv \delta g_{\nu_\tau}$,
and $1/x$ for a range of values of mixing angle $\phi$.  At each value
of $\phi$ we fixed the coupling shifts to their best-fit values and
used the calculated one-sigma error on $1/x$ to determine a minimum
allowed mass for the $Z'$ and $W'$ at 95\% c.l.  The resulting
exclusion curves are shown in Figures \ref{fig:hncetc} and
\ref{fig:lncetc}.

In heavy NCETC, the extra weak bosons are allowed to be lightest  when the
mixing angle is at a value near 
$\sin^2\phi = 0.75$.  At this point, the best-fit values for the other
model parameters are
\begin{eqnarray}
\delta g_b &=& -0.0015 \pm 0.00089 \nonumber \\
\delta g_\tau &=& -0.00061 \pm 0.00066 \\
1/x &=& 0.00166 \pm 0.0011 \nonumber
\label{eqn:hncetcval}
\end{eqnarray}
The corresponding best-fit value for the Z' mass is 4.55 TeV, while the
minimum allowed value of the Z' mass at 95\% c.l. is 2.98 TeV.  The
goodness of fit is 6.6\%, as compared with 7.3\% when we fit the SM
predictions to the same set of data.

In light NCETC, the Z' and W' can be least massive when the
mixing angle is at a value near 
$\sin^2\phi = 0.60$.  At this point, the best-fit values for the other
model parameters are
\begin{eqnarray}
\delta g_b &=& -0.00080 \pm 0.00085 \nonumber \\
\delta g_\tau &=& -0.00017 \pm 0.00076 \\
1/x &=& -0.0013 \pm 0.0031 \nonumber
\end{eqnarray}
Note that the best-fit value of 1/x corresponds to an unphysical value
for the Z' or W' mass. We therefore take the maximum (positive) value
of 1/x at the 2$\sigma$ level ({\it i.e.} 1/x=+0.00491) and use the
relation $M^H_W \approx {\sqrt{x}\over s c} M^0_W$ to determine the
minimum $M_{W'}$ value of 2.3 TeV. Fixing 1/x at this maximum positive
value, we performed a two-parameter fit to the couplings, obtaining
best-fit values of
\begin{eqnarray}
\delta g_b &=& 0.00039 \pm 0.00061 \nonumber \\
\delta g_\tau &=& 0.0010 \pm 0.00048 
\end{eqnarray}
The goodness of fit is 2.9\%, as compared with 5.6\% when we fit the
SM predictions to the same set of data.

At energies well below the mass of the NCETC Z' boson, its exchange in the
process $e^+ e^- \to f \bar{f}$ where $f$ is a $\tau$ lepton or
b-quark may be approximated by the contact interaction
\begin{equation} 
{\cal{L}}_{NC} \supset \frac{e^2}{\sin^2\theta\, M^2_{Z'}}
\left(-\frac{c_\phi}{2 s_\phi} \big(\bar{e_L} \gamma_\mu
  e_L\big)\right) \left(\frac{s_\phi}{2 c_\phi} 
  \big(\bar{f}_L \gamma^\mu f_L\big)\right)\ , 
\end{equation}
based on the Z'-fermion couplings in eqn.(\ref{eqn:ncetc-coupl}).
As noted earlier, this Z' couples only to left-handed fermions to first
approximation.  Comparing this with the contact interactions studied
by LEP (\ref{lepcontact}), we find
\begin{equation}
M_{Z'} = \Lambda^+ \sqrt{\frac{\alpha_{em}}{4\sin^2\theta }}\ .
\label{zplimencetc}
\end{equation}
The LEP limits from tau lepton production ($\Lambda^+_{LL} \geq 11.4$
TeV) and from b-quark production ($\Lambda^+_{LL} \geq 11.8$) TeV are
comparable.  Using eqn. (\ref{zplimencetc}), we find $M_{Z'} \geq 1.1$
TeV, a weaker bound than that provided by the precision electroweak
data.  Again, hadron collider limits, being suppressed by $\cos\phi /
\sin\phi$ are even weaker.

Neutral B-meson mixing has also been used \cite{Simmons:2001va} to set
lower bounds on the Z' mass in these models.  If all quark mixing is
assumed to occur in the left-handed down sector, a lower limit of order a
TeV results -- far weaker than the limits we have set using the precision
electroweak data.  The B-mixing limit is also, as noted in
ref. \cite{Simmons:2001va}, highly dependent on the flavor structure
assumed for the model.

\section{Weak Bosons in the Ununified Standard Model}
\label{sec:ununified}
\setcounter{equation}{0}

\begin{figure}
\begin{center}
{\rotatebox{0}{\scalebox{1.0}{\includegraphics{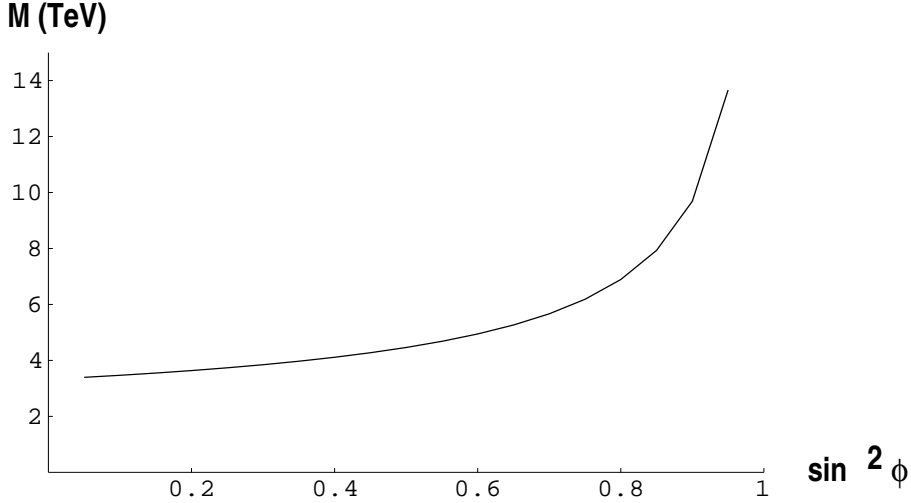}}}}
\end{center}
\caption[down]{\small Lower bound on UUM Z' boson mass at 95\% CL 
as a function of
mixing angle.}
\label{fig:uum}
\end{figure}

As described in
ref. \cite{Georgi:1989ic,Georgi:1989xz,Chivukula:1994qw}, this model
is based on the electroweak gauge group $SU(2)_q \times SU(2)_\ell
\times U(1)$.  Left-handed quarks and leptons transform as doublets
under $SU(2)_q$ and $SU(2)_\ell$, respectively; right-handed quarks
and leptons transform as singlets\footnote{See
\protect\cite{Georgi:1989xz} for comments on the use of additional
fermions to cancel the $SU(2)_q^{\,2}\times U(1)$ and
$SU(2)_\ell^{\,2} \times U(1)$ anomalies.} under both $SU(2)$ gauge
groups.  The $U(1)$ is the hypercharge group of the standard
model. The gauge couplings may be written
\beq \gq={e\over \sp\st}\,,\quad \gl={e\over \cp\st}\,,\quad
\gp={e\over \ct}\,, 
\eeq 
in terms of the usual weak mixing angle
$\theta_W$ and a new mixing angle $\phi$.

The electroweak gauge group
spontaneously breaks to $U(1)_{em}$ which is generated by
$Q= T_{3q}+T_{3\ell}+Y.$  This symmetry breaking occurs
when two scalar fields, $\Phi$
and $\Sigma$,  transforming respectively as
$(1,2)_{1/2}$ and $(2,2)_0$ acquire the vacuum expectation values (vev's)
\beq
\left\langle\Phi\right\rangle=\pmatrix{ 0\cr {v/{\sqrt
2}}\cr}\,,\quad\left\langle\Sigma\right\rangle=\pmatrix {u&0\cr
0&u\cr}\,\,\,.
\eeq
The vev of $\Sigma$ breaks the two $SU(2)$'s down to the diagonal
$SU(2)_W$ of the standard model.   Thus this theory reproduces the
phenomenology of the standard model for $u\gg v$. 

In the limit of large $x \equiv u^2/v^2$, the light gauge boson mass 
eigenstates are 
\beq
W^L\approx W_1+{s^3c\over x}\,W_2\,,\quad  
Z^L\approx Z_1+{s^3c\over x\,\cos\theta}\,Z_2\,\,.
\eeq
and they couple to fermions as, respectively,  
\beqa
{e\over\sin\theta}\left( T_q^\pm + T_\ell^\pm  + {s^2 \over x} {\left(
 c^2 T_q^\pm - s^2 T_\ell^\pm \right) } \right) \nonumber \\
{e \over {\st \ct}} \left( T_{3q} + T_{3\ell} - \sin^2
 \theta \, Q +{s^2 \over x} {\left( c^2 T_{3q} - s^2 T_{3\ell}
\right) } \right) \, .
\label{eqn:uum-coupl}
\eeqa 
In this approximation, the heavy eigenstates have a mass given
by ${M^H_W\over M^0_W}\approx {M^H_Z\over
M^0_Z}\approx {\sqrt{x}\over s c} .$
where $M^0_W$ is the tree-level W-boson mass in the Standard Model.

The presence of the extra W and Z bosons in this model leads to
predicted deviations in the values of electroweak observables.  The
list of affected quantities is indicated in the last column of Table
\ref{tab:obs-sm}.  Note that while quarks and leptons couple
differently to the electroweak bosons in the model, generation
universality is preserved so that the ``leptonic'' values of $A_{FB}$
and $R$ are relevant rather than the distinct values measured for each
lepton species; likewise it is the value of $\sigma_h$ which assumes
lepton universality which is relevant here.  The formulas for the
predicted shifts from SM predictions are discussed in Appendix C.

We performed a global fit
of the electroweak data to the model's predictions and determined a 95\%
c.l. lower bound on $M_{Z'}$ as a function of the mixing angle $\phi$, as
shown in Figure \ref{fig:uum}.  The mass of the heavy Z' and W' states must
always be at least 3.4 TeV, with the limit being stronger as $\sin\phi$
increases.  The quality of fit for the Ununified Model on the limit curve
is 2.9\%, as compared with 4.1\% when we fit the predictions of the SM to
the same data.

In this model, limits on contact interactions tend to provide much
weaker bounds on the Z' mass than the precision electroweak data.  The
strongest limits from contact interactions arise from the process
$e^+_L e^-_L \to \bar{b}_L b_L$, for which (based on
eqn. (\ref{eqn:uum-coupl})) 
\begin{equation}
M_{Z'} = \Lambda^{+} \sqrt{\frac{\alpha_{em}}{4\sin^2\theta }}.
\label{zplimeuum}
\end{equation}
LEP finds \cite{Abbaneo:2002xx} $\Lambda^+_{LL} \geq 11.8$ TeV, implying
$M_{Z'} \geq 1.1$ TeV.  The $M_{Z'}$ limit from $e^+_L e^-_L \to
\ell^+_L \ell^-_L$, for which LEP finds $\Lambda^-_{LL} \geq 9.8$ TeV,
is suppressed by a factor of $\sin\phi / \cos\phi$ because only
leptons are involved.  Tevatron limits on quark compositeness have the
potential to be stronger because they are enhanced by a factor of
$\cos\phi / \sin\phi$; but the existing D0 bound \cite{Abbott:1997nf} 
$\Lambda^- \geq 2.2$ TeV implies only $M_{Z'} \geq 900$ GeV even when
$\sin^2\phi = 0.05$.

\section{Conclusions}

In this note we update the bounds
\cite{Chivukula:1996cc,Chivukula:1996gu,Chivukula:1994qw} placed by
electroweak data on the existence of flavor non-universal extensions
to the standard model in the context of topcolor assisted technicolor
(TC2), noncommuting extended technicolor (NCETC), and the ununified
standard model (UUM). 

We find that the the extra $Z$ in TC2 models must be heavier than
about 2 TeV for generic values of the gauge coupling.  However, for
values of the new gauge boson mixing angle near $\sin\phi \approx
0.0784$, cancellations among parameters limit the size of deviations
of Z-pole observables, weakening the precision electroweak limits.  In
this region of parameter space, a stronger lower bound on the $Z'$
mass comes from limits on contact interactions at LEP II, which imply
that the TC2 $Z'$ must be greater than about 1 TeV.  For TC2 models,
limits on the $Z'$ mass from flavor-changing neutral currents have
been found to be quite model-dependent, in contrast with the limits
reported here.  We note that a lower bound of order a TeV on the TC2
$Z'$ mass is consistent with the goal of providing sufficient
dynamical electroweak symmetry breaking without fine-tuning
\cite{Hill:1994hp}.

The extra $SU(2)$ triplet of gauge bosons in NCETC and UUM models must
be somewhat heavier, with masses always greater than about 3 TeV. The
limits on these models from Z-pole observables are significantly
stronger than those from contact interactions at LEP II or from
flavor-changing neutral currents.  In the context of NCETC, a lower
bound of order 3 TeV on the masses of the extra SU(2) gauge-bosons
implies that the scale of the ETC interactions responsible for
generating the top-quark mass must also be greater than about 3
TeV. As noted in \cite{Chivukula:1996gu}, this implies that
the ETC interactions must be strongly-coupled and that fine-tuning is
required in order to accommodate a top-quark mass of 175 GeV. Using the
estimates in \cite{Chivukula:1996gu}, we see that the strong ETC coupling
must be adjusted to of order a few percent or less.

\centerline{\bf Acknowledgments}

We thank the referee for carefully reading the manuscript.  {\em This
work was supported in part by the Department of Energy under grant
DE-FG02-91ER40676 and by the National Science Foundation under grant
PHY-0074274.}


\section*{Appendix A: Corrections for TC2}
\renewcommand{\theequation}{A.\arabic{equation}}
\label{sec:tc2appx}
\setcounter{equation}{0}

The full list of electroweak corrections to standard
model  predictions in TC2 is:
\begin{equation}
 \Gamma_Z = \left( \Gamma_Z \right)_{SM} 
\left( 1 +\left[ - 0.0390 \ttphi + 0.0520 \ssphi + 0.00830 \ccphi \right]
{1 \over x} \right)
\end{equation}
\begin{equation}
 A_{LR} = \left( A_{LR} \right)_{SM}  + 
\left[ 1.986 \ttphi - 0.202 \ssphi + 0.00366 \ccphi \right]
{1 \over x}
\end{equation}
\begin{equation}
 A_{FB}^e = \left( A_{FB}^e \right)_{SM}  + 
\left[ 0.474 \ttphi - 0.483 \ssphi + 0.000875 \ccphi \right] 
{1 \over x}
\end{equation}
\begin{equation}
 A_{FB}^\mu = \left( A_{FB}^\mu \right)_{SM}  + 
\left[ 0.474 \ttphi - 0.483 \ssphi + 0.000875 \ccphi \right] 
{1 \over x}
\end{equation}
\begin{equation}
 A_{FB}^\tau = \left( A_{FB}^\tau \right)_{SM} + 
\left[ 0.474 \ttphi - 0.214 \ssphi + 0.0139 \ccphi \right]
{1 \over x}
\end{equation}
\begin{equation}
 \sigma_h = \left( \sigma_h \right)_{SM} 
\left(1 - \left[0.0152 \ttphi - 0.105 \ssphi + 0.00830 \ccphi \right]
{1 \over x} \right)
\end{equation}
\begin{equation}
 R_b = \left( R_b \right)_{SM} 
\left(1 - \left[0.0440 + 0.190 \ssphi - 0.0146 \ccphi \right]
{1 \over x} \right)
\end{equation}
\begin{equation}
 R_c = \left( R_c \right)_{SM} 
\left(1 - \left[0.0944 - 0.0625 \ssphi + 0.00432 \ccphi \right]
{1 \over x} \right)
\end{equation}
\begin{equation}
 R_e = \left( R_e \right)_{SM}
\left( 1 + \left[ .200 \ttphi + 0.0325 \ssphi - 0.00378 \ccphi \right]
{1 \over x} \right)
\end{equation}
\begin{equation}
 R_\mu = \left( R_\mu \right)_{SM} 
\left(1 + \left[0.200 \ttphi + 0.0325 \ssphi - 0.00378 \ccphi  \right]
{1 \over x} \right)
\end{equation}
\begin{equation}
 R_\tau =
 \left( R_\tau \right)_{SM} 
\left(1 + \left[0.200 \ttphi - -.316 \ssphi + 0.0235 \ccphi \right] 
{1 \over x} \right)
\end{equation}
\begin{equation}
 A_{e}(P_\tau) = \left( A_{e}(P_\tau) \right)_{SM}  + 
\left[ 1.986 \ttphi - 0.202 \ssphi + 0.00366 \ccphi \right] 
{1 \over x}
\end{equation}
\begin{equation}
 A_{\tau}(P_\tau) = \left( A_{\tau}(P_\tau) \right)_{SM} +
\left[ 1.986 \ttphi - 1.592 \ssphi + 0.113 \ccphi \right] 
{1 \over x}
\end{equation}
\begin{equation}
 A_{FB}^b = \left( A_{FB}^b \right)_{SM} +
\left[ 1.414 \ttphi - 0.157 \ssphi + 0.00365 \ccphi \right]
{1 \over x}
\end{equation}
\begin{equation}
 A_{FB}^c = \left( A_{FB}^c \right)_{SM}  + 
\left[ 1.105 \ttphi - 0.113 \ssphi + 0.00204 \ccphi \right]
{1 \over x}
\end{equation}
\begin{equation}
 {\cal A}_b = \left( {\cal A}_b \right)_{SM} +
\left[ 0.161 \ttphi - 0.129 \ssphi + 0.00912 \ccphi \right]
{1 \over x}
\end{equation}
\begin{equation}
 {\cal A}_c = \left( {\cal A}_c \right)_{SM}  + 
\left[ 0.867 \ttphi - 0.0883 \ssphi + 0.00160 \ccphi \right]
{1 \over x}
\end{equation}
\begin{equation}
 M_W = \left( M_W \right)_{SM} 
\left(1 - \left[0.165 \ttphi - 0.0258 \ssphi + 0.00101 \ccphi\right] 
{1 \over x} \right)
\end{equation}
\begin{equation}
 g_L^2(\nu N \rightarrow \nu X) =
\left( g_L^2\right)_{SM} + 
\left[ 0.0576 \ttphi - 0.0194 \ssphi + 0.00121 \ccphi \right] 
{1 \over x}
\end{equation}
\begin{equation}
 g_R^2(\nu N \rightarrow \nu X) =
\left( g_R^2\right)_{SM} +
\left[ -0.0196 \ttphi + 0.00666 \ssphi - 0.0000350 \ccphi \right]  
{1 \over x}
\end{equation}
\begin{equation}
 Q_W(Cs) = \left( Q_W(Cs) \right)_{SM}  + 
\left[ 16.57 \ttphi - 5.655 \ssphi - 0.00206 \ccphi\right]
{1 \over x}
\end{equation}

\section*{Appendix B: Corrections for NCETC}
\renewcommand{\theequation}{B.\arabic{equation}}
\label{sec:ncetcappx}
\setcounter{equation}{0}

The formulae for corrections to most of the variables used in our fits
are given in \cite{Chivukula:1996gu}.  Those for the few additional
variables used here are below.

\bigskip
\noindent{\tt Heavy Case}

\begin{equation}
 R_c = \left( R_c \right)_{SM} 
\left(1 - 1.01(\delta g^b_L)^{ETC} + 
\left[0.505 s^4 + 1.40 s^2c^2 - 0.121 (1-s^4) \right]
{1 \over x} \right)
\end{equation}
\begin{equation}
 {\cal A}_b = \left( {\cal A}_b \right)_{SM} - 0.293(\delta g^b_L)^{ETC} +
\left[ -0.146 s^4 - 0.208 (1-s^4) \right]
{1 \over x}
\end{equation}
\begin{equation}
 {\cal A}_c = \left( {\cal A}_c \right)_{SM}  + 
\left[ -0.785 s^2c^2 - 1.123 (1-s^4)  \right]
{1 \over x}
\end{equation}
\begin{equation}
 A_e = 1-{2\over x}
\end{equation}

\noindent{\tt Light Case}
\begin{equation}
 R_c = \left( R_c \right)_{SM} 
\left(1 - 1.01(\delta g^b_L)^{ETC} + \left[- 1.01 s^2c^2 - 1.784c^4 \right]
{1 \over x} \right)
\end{equation}
\begin{equation}
 {\cal A}_b = \left( {\cal A}_b \right)_{SM} - 0.293(\delta g^b_L)^{ETC} +
\left[ 0.146 s^2c^2 + 0.208 c^4 \right]
{1 \over x}
\end{equation}
\begin{equation}
 {\cal A}_c = \left( {\cal A}_c \right)_{SM}  + 
\left[ 1.908 c^4  \right]
{1 \over x}
\end{equation}

\section*{Appendix C: Corrections for the UUM}
\renewcommand{\theequation}{C.\arabic{equation}}
\label{sec:uumappx}
\setcounter{equation}{0}

The formulae for corrections to most of the variables used in our fits
are given in \cite{Chivukula:1994qw}.  Those for the few additional
variables used here are below.

\begin{equation}
 R_c = \left( R_c \right)_{SM} 
\left(1 + \left[0.073 s^2c^2 + 0.121 s^4 \right]
{1 \over x} \right)
\end{equation}
\begin{equation}
 {\cal A}_b = \left( {\cal A}_b \right)_{SM} +
\left[ 0.1467 s^2c^2 + 0.208 s^4 \right]
{1 \over x}
\end{equation}
\begin{equation}
 {\cal A}_c = \left( {\cal A}_c \right)_{SM}  + 
\left[ 0.7855 s^2c^2 + 1.123 s^4  \right]
{1 \over x}
\end{equation}


\end{document}